# Title page

**Title**

A 60-GHz Radar Sensor for Micron-Scale motion detections

**Short running title**

60-GHz Radar Sensor

**Authors**

Marcel Balle[1], Chengkai Zhu[1], Bin Zhang[1], Jie Wang[2] and Lixin Ran[1]

**Affiliation 1**

Laboratory of Applied Research on Electromagnetics (ARE), Zhejiang University, Hangzhou 310027, China.

**Affiliation 2**

Laboratory of Antenna Feed System, Beijing Institute of Remote Sensing Equipment, Beijing 100854, China.

**Corresponding authors**

Jie Wang (wangjie_soph@126.com) and Lixin Ran (ranlx@zju.edu.cn)

# A 60-GHz Radar Sensor for Micron-Scale Motion Detection

**Abstract**

A compact, continuous-wave, mmWave radar sensor is developed for non-contact detection of micron-scale motions. This board-integrated radar system consists of a pair of mmWave transmitter and receiver, two series-fed microstrip patch arrays, an IF subsystem, and a microcontroller. Working at 60-GHz frequency, this super-heterodyne, digital-IF radar sensor exhibits an agile sensitivity and a robust anti-noise performance. Assisted by a gradient-descent DC offset estimation and a phase demodulation algorithm based on extended differential and cross-multiply, a 45-μm pendulum swing can be experimentally detected with a 20.6-dB SNR, verifying its strong ability in Doppler micro-motion detections.

**1. Introduction**

Recently, there are more and more demands for non-contact monitoring of micro-scale motions. Typical examples include mechanic vibrations in mechanic systems such as ultrasound transducers, biomotions of human heartbeats and wrist pulses conducted to the skin [1]-[3]. It is therefore essential to develop non-intrusive monitoring systems capable of detecting such motions. Doppler radar sensors (DRSs) are able to present advantageous non-obstructive illumination, insensitivity to lighting conditions and privacy preservation. Developing high-performance DRSs is thus a promising solution.

Recently, motion detections based on mmWave DRSs have been attracting more and more interests. With the advancement of mmWave integrated circuits (MMIC), many interests have focused on the development of miniaturized, low-power and low-cost DRSs for detecting both small- and large-scale biomotions, especially human vital signs detections [1]-[4]. Most of these studies utilized DRSs operating at microwave frequencies. However, monitoring micro-scale biomotions with centimeter wavelengths is challenging. On the other hand, adopting mmWave frequencies with much shorter wavelengths will significantly increase the sensitivity and precision of the DRS. Furthermore, in order to avoid the DC offset and flicker noise overshadowing low-end frequencies, the digital-IF architecture is preferable for DRSs aiming to detect low-frequency biomotions.

In this paper, a compact, continuous-wave, mmWave DRS is designed and implemented. This board integrated mmWave system consists of a pair of mmWave transmitter and receiver, two series-fed microstrip patch arrays, an IF subsystem, and a microcontroller. Working at 60-GHz frequency, this super-heterodyne, digital-IF radar sensor exhibits an agile sensitivity and a robust anti-noise performance for detecting micron-scale motions. Experimental measurements to the micron-scale swing of a single pendulum verified its performance in detecting micron-scale motions.

## 2. Algorithms

In a typical DRS, the transmitting antenna radiates an electromagnetic wave to illuminate a moving target. On the principle of the Doppler effect, upon reflections on the target with time-varying motion *x(t)*, at an initial detection distance $d_0$ from the antennas, the signal will be phase shifted. According to [5], for typical quadrature transceivers, the demodulated baseband in-phase (*I*) and quadrature-phase (*Q*) signals can be described as

$$I(t) = A_I(t) \cos\left[\theta_0 + \frac{4\pi x(t)}{\lambda} + \Delta\varphi(t)\right] + DC_I(t) \tag{1}$$

$$Q(t) = A_Q(t) \sin\left[\theta_0 + \frac{4\pi x(t)}{\lambda} + \Delta\varphi(t)\right] + DC_I(t) \tag{2}$$

where $A_I$ and $A_Q$ denote the amplitude of the *I/Q* signals. For DRSs using a digital-IF architecture, the imbalance between $A_I$ and $A_Q$ can be compensated [2]. $\theta_0$ is a constant phase delay due to the initial distance, which can be suppressed by calibration. $\lambda$ is the signal wavelength and $\Delta\varphi(t)$ represents the residual phase noise that can be neglected by the range correlation effect [6]. $DC_I(t)$, $DC_Q(t)$ denote the DC offsets caused by the stationary background scattering. From (1) and (2), motion *x(t)* can be linearly retrieved with the direct arctangent-based phase demodulation

$$\Phi(t) = \frac{4\pi x(t)}{\lambda} = arctan\left[\frac{Q(t)-DC_Q(t)}{I(t)-DC_I(t)}\right] \tag{3}$$

In practice, to avoid the inherent phase unwrapping issue of the arctangent function, (3) can be replaced by an equivalent extended differentiate and cross-multiply (DACM) algorithm [1], whose discrete form is

$$x[n] = \frac{\lambda}{4\pi}\sum_{k=2}^{n}\frac{I[k]\{Q[k]-Q[k-1]\}-Q[k]\{I[k]-I[k-1]\}}{I^2[k]+Q^2[k]} \tag{4}$$

According to (3), the key to obtain a precise *x(t)* is to know the accurate DC offset. Based on (1) and (2),

$$[I(t) - DC_I(t)]^2 + [Q(t) - DC_Q(t)]^2 = A^2(t) \tag{5}$$

It is apparent that the detected *I/Q* signals are located on a circle where the balanced amplitude *A(t)* is the radius and the Cartesian coordinate [$DC_I(t)$, $DC_Q(t)$] is the center. In practice, the circle radius and center can be optimized based on detected *I/Q* signals. An example is the gradient-descent (GD) algorithm proposed in [7].

## 3. Implementation of the mmWave DRS

Illustrated in Fig. 1 is the block diagram of the proposed DRS based on the 60-GHz MMIC transmitter and receiver, i.e., ADI[TM] HMC6300 and HMC6301, and the digital-IF subsystem, i.e., ADI[TM] AD9874. The IF frequency was chosen as 250 MHz, and the bandpass sampling was used to obtain the digitized baseband *I/Q* signals. Finally, a micro-controller unit (MCU), ST

microelectronics™ STM32F103RET6, was used to configure the radar system and provide a series communication with a computer.

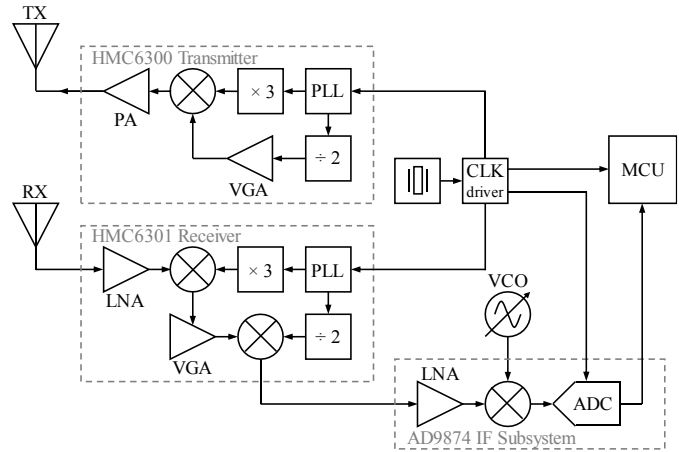

Fig. 1. Functional block diagram of the 60-GHz digital-IF DRS.

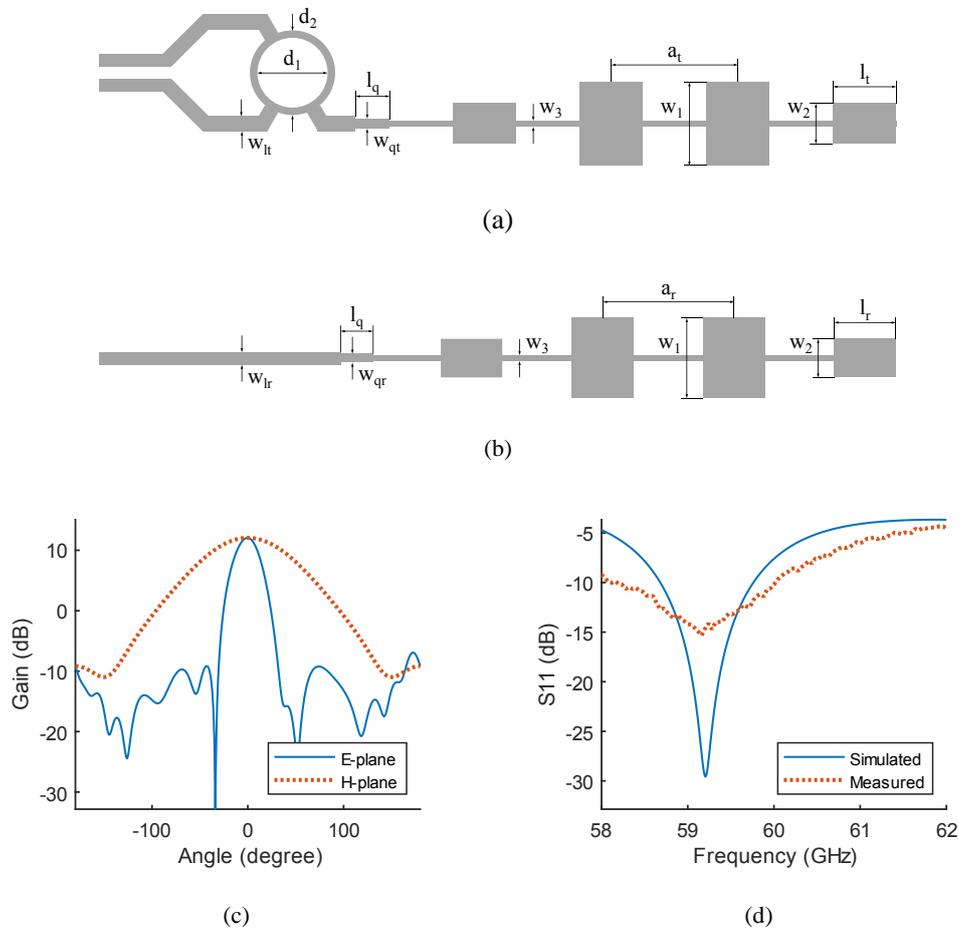

Fig. 2. Design of the antennas. (a) Transmitting antenna designed with a microstrip balun. $w_1$ = 1.669 mm, $l_t$ = 1.215 mm, $a_t$ = 2.51 mm, $w_2$ = 0.806 mm, $w_3$ = 0.12 mm, $l_q$ = 0.67 mm, $w_{qt}$ = 0.196 mm, $w_{lt}$ = 0.3 mm, $d_1$ = 1.39 mm, $d_2$ = 1.69 mm (b) Receiving antenna. $l_r$ = 1.274 mm, $a_r$ = 2.7 mm, $w_{qr}$ = 0.179 mm, $w_{lr}$ = 0.254 mm (c) Simulated radiation patterns. (d) Simulated and experimental reflection coefficients.

For the 60-GHz transmitting and receiving MMICs, the accuracy design of the antennas is the key to ensure the performance of the DRS. Figs. 2(a) and 2(b) show the structure and dimensions of the series-fed 4-element patch arrays designed for the transmitter and receiver, respectively. The substrate was chosen as the 0.167-mm-thick Rogers$^{TM}$ RO4350B, whose dielectric constant is 3.48. A microstrip balun is specially designed to satisfy the requirement of the transmitter MMIC. For effective reduction of the elevation beam sidelobes, the Chebyshev synthesis was used to obtain the excitation coefficients and taper each patch element accordingly.

Fig. 2(c) shows the radiation pattern simulated for the E- and H-planes. The simulated antenna gain is about 12 dBi, with 3-dB beamwidths around 25.4° and 86°. The sidelobe levels are -19 dB and -21.2 dB in the E- and H-planes, respectively. Both the simulated and the measured reflection coefficients (S11s) are plotted in Fig. 2(d). It can be seen that both curves have a maximum radiation at 59.2 GHz. Although the measured 15-dB S11 is worse than the simulated one, it is satisfactory in practical applications.

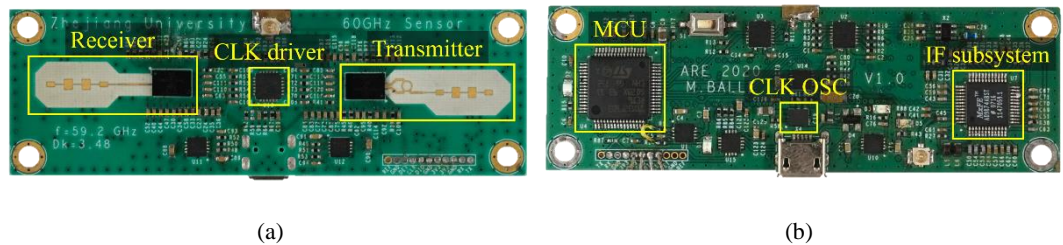

(a)            (b)

Fig. 3. Photographs of the 80 ×26 mm$^2$ board-integrated 60-GHz DRS. (a) Top view. (b) Bottom view.

**4. Experimental verification**

In order to determine the sensitivity and linearity of the proposed DRS as well as the effectiveness and accuracy of the signal processing algorithms, the sinusoidal motion of a simple pendulum is measured. Fig. 4(a) describes the experimental setup placed in an anechoic chamber. The constant frequency of the pendulum oscillation depends on the arm length $L$ between the pivot point and center of the pendulum bob, as expressed by $f = \sqrt{g/L}/(2\pi)$, where $g$ is the gravitational acceleration.

For the experiment, $L$ was adjusted to approximately 6 cm, corresponding to an oscillation frequency of 2.04 Hz. With such a short arm length, the pendulum amplitude will dampen rapidly so that the motion scale recorded during one measure sequence will vary significantly. The DRS was placed about 30-cm away from the equilibrium position of the pendulum. Then, the reflected signal modulated by the weak sinusoidal motion was recorded by the DRS for 120 seconds with a sampling frequency of 100 Hz. The constellation diagram of the sampled *I/Q* data, along with the circle fitted by the DC offset estimation algorithm, are illustrated in Fig. 4(b).

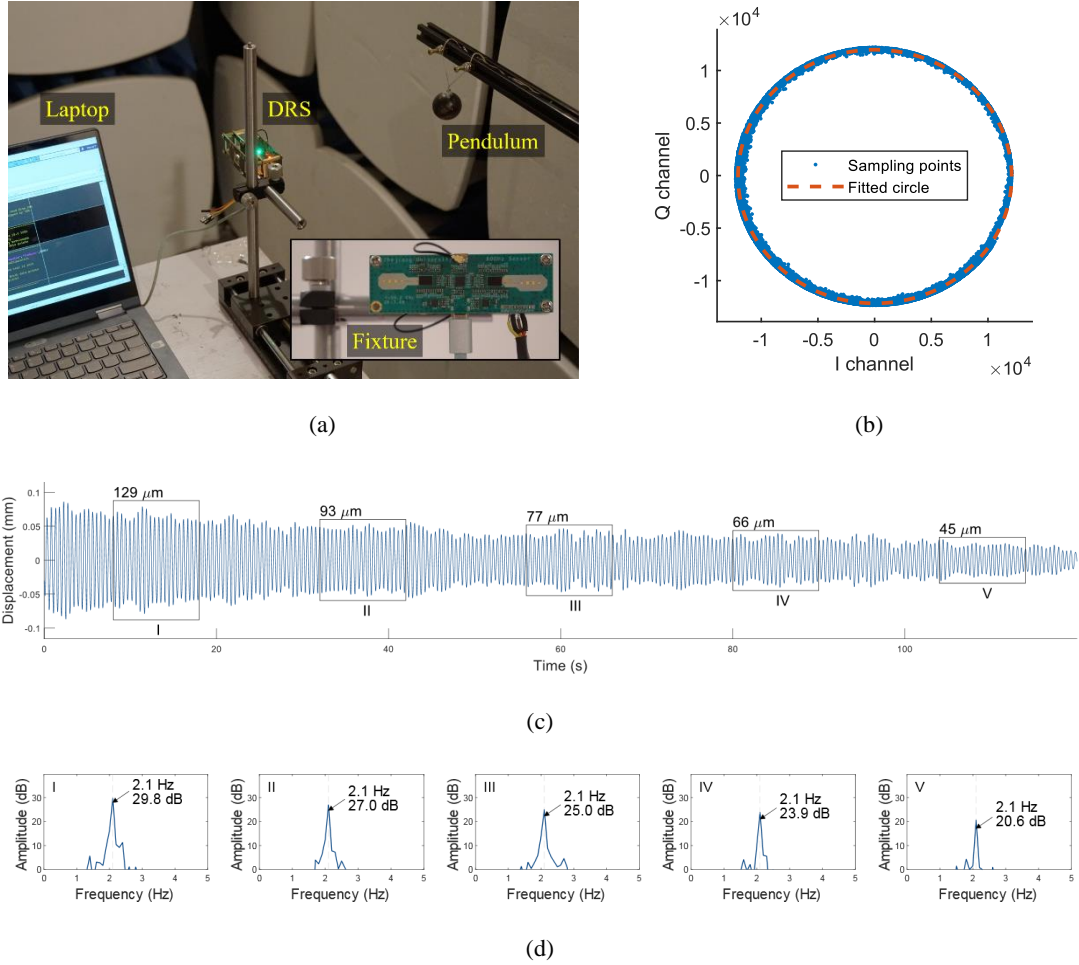

Fig. 4. Pendulum experiment. (a) Experimental setup. (b) Constellation diagram and fitted circle. (c) DACM-reconstructed pendulum motion. (d) Spectra of five segments.

Fig. 4(c) shows the 120-seconds pendulum swing reconstructed by the gradient-descent DC offset estimation combined with the extended DACM algorithm. The spectra of five segments chosen from the 120-second data are shown in Fig. 4(d). It is observed that due to the ultra-small amplitude, a quasi-sinusoidal motion with its swing amplitude decaying from ~129 μm to ~45 μm can be accurately recovered. Even when the pendulum swing is 45 μm, the corresponding SNR is still larger than 20 dB. The 2.1-Hz oscillation frequency computed by the fast Fourier transform (FFT) for each segment roughly agrees with the previously calculated value. Furthermore, no visible harmonics can be observed in the spectra, suggesting a high linearity of the DRS and the signal processing algorithms.

## 5. Conclusion

In this work, a highly integrated 60-GHz continuous-wave DRS implemented with high-performance super-heterodyne and digital-IF architectures and on-board printed microstrip patch arrays is designed, simulated and experimentally implemented. Experimental results suggested that the developed sensor can be effectively used to detect micro-scale motions with satisfactory SNRs. Because of its compact size and low power consumption, this board-integrated radar sensor is well suited for monitoring weak biomotions such as human cardiac motions, wrist pulses and blood pressures. We envision its wide applications in non-intrusive daily healthcare and micro-vibration monitoring in a near future.

**Acknowledgements**

This work was supported by the National Natural Science Foundation of China under grants 61771421, 62071417 and 62111530096.